\documentclass[pra,aps,showpacs, twocolumn]{revtex4}
\usepackage{mathtext}
\usepackage[T2A]{fontenc}
\usepackage[cp1251]{inputenc}
\usepackage{epsf}
\usepackage{psfrag}
\usepackage{graphicx}
\usepackage{amssymb,amsmath}

\usepackage{color}

\begin{document}

\title{Influence of atomic motion on the collective effects in dense and cold atomic ensembles}
\author{A. S. Kuraptsev and I. M. Sokolov
\\
%EndAName
{\small Peter the Great St. Petersburg Polytechnic University, 195251, St. Petersburg, Russia}\\
{\small ims@is12093.spb.edu}\\
}
\date{\today}

\sloppy

%\affiliation

%\baselineskip22 pt\newpage

\begin{abstract}
We show that atomic motion leads not only to noticeable quantitative, but in some cases also to qualitative modification of collective effects in dense and cold atomic ensembles even in the case when the characteristic Doppler shifts are tens of times smaller than the natural linewidth. The observed influence is explained as a result of the suppression of the impact of sub-radiant collective states caused by the displacement of the atoms.

\end{abstract}


\maketitle

Collective effects such as sub and superradiance, weak and strong (Anderson) localization have recently attracted keen interest. Atomic motion can significantly modify the character of these effects. The light emitted by one atom turns out to be non-resonant to the transition of another one moving with a different speed. In the case when the Doppler shifts are greater or comparable with the natural width of the atomic excited state, the mutual non-resonance of the atoms is taken into account by introducing a random shifts of atomic levels that are different for different atoms (\cite{JRSY14,JRJBPSB16}).

Lowering the temperature weakens the Doppler effect, therefore one of the most interesting objects for the study of collective effects is the atomic ensembles, cooled to sub-Doppler temperatures in special traps \cite{Araujo:2016, Roof:2016, Bromley:2016, Jennewein:2018}. Thus, in modern magneto-optical and dipole traps, atomic clouds are cooled to temperatures of the order of 30-100 $\mu K$. In this case, the Doppler shifts and the inhomogeneous broadening of the lines are substantially less than the natural linewidth and the dipole-dipole interaction is practically indistinguishable from the case of fixed atoms (see for example \cite{Stephen_1964, JRSY14}). Even the frequency diffusion due to the possible large number of light scattering by atoms is usually irrelevant \cite{Skipetrov_2016}. For this reason, when describing collective effects the model of motionless scatterers is usually used. The displacement of atoms due to the low but finite temperature is taken into account by averaging the observed values over the random spatial configurations of atoms in the ensemble.

In this paper, we show that such approach can lead not only to substantial quantitative errors, but also in some cases to qualitatively wrong conclusions, even when the Doppler frequency shifts are few tens times smaller than the natural width of the atomic transition. As an illustration, we consider two problems. In the first, we calculate the transmission of the atomic cloud and show that the time average transmittance obtained as a solution of the dynamic problem for continuously transient atoms may significantly differ from the results obtained by averaging over the random spatial configurations of immobile atoms with the same spatial distribution. The second example deals with the dynamics of the spontaneous decay in a cloud of slowly moving atoms. For times exceeding the natural lifetime, a significant effect of motion on decay rate is also found out.

We consider an ensemble consisting of $ N >> 1 $ identical atoms with a nondegenerate ground state with an angular momentum $ J_g = 0 $. The excited  state is $ J_e = 1 $. The lifetimes of all three of its Zeeman sublevels ($ m  = -1,0,1 $) are the same and equal to $ 1 / \gamma $.
The evolution of the atomic states is described by means of the coupled dipoles  model, which is traditional for this class of the problems.
This model was first proposed by Foldy \cite{F45}, then discussed in detail by Lax \cite{L51}. Later similar approach was used in the context of different type of collective effects such as multiple and recurrent scattering, collective spontaneous decay and Anderson localization of light \cite{Javanainen:1999}-\cite{Skipetrov_2019}.

We analyse the properties of a closed system consisting of all atoms and an electromagnetic field, including a vacuum reservoir. We look for the wave function $ \psi $ of this system in the form of expansion over the eigenfunctions $ {\psi_l} $ of the Hamiltonian of noninteracting atoms and light $ \psi = \sum_l b_l \psi_l $. Considering the case of weak excitation and restricting ourselves by the states of the atomic-field system containing no more than one photon, for the amplitudes $ b_e $ of one-fold excited atomic states $ \psi_e = | g \cdots e \cdots g \rangle $ we have the following set of equations

\begin{equation}
\frac{\partial b_e}{\partial t} = \left( i\delta_e-\frac{\gamma}{2} \right)b_e -\frac{i\Omega_{e}}{2} + \frac{i\gamma}{2} \sum_{e' \neq e} V_{ee'}b_{e'}.
\label{e1}
\end{equation}
Here, the index $ e $ shows the number of the atom which is excited in the state $ \psi_e = | g \cdots e \cdots g \rangle $, as well as specific populated Zeeman sublevel.

The first term on the right side of Eq. (\ref{e1}) corresponds to the natural evolution of independent atomic dipoles. The second one describes the influence of the external laser field. The Rabi frequency of the field in the point where atom $e$ is located is $\Omega_e$. Its detuning $\delta_e$ may be different for different transitions $g\leftrightarrow e$. Such difference takes place in the presence of external static electric or magnetic field. The last term in Eq. (\ref{e1}) corresponds to the dipole-dipole interaction and is responsible for all collective effects.  It reads
\begin{eqnarray}
V_{ee'}& =&
-\frac{2}{\gamma} \sum\limits_{\mu, \nu}
\mathbf{d}_{e g}^{\mu} \mathbf{d}_{g e'}^{\nu}
\frac{e^{i k_0 r_{ij}}}{\hbar r_{ij}^3}
\nonumber
\\
&\times& \left\{
\vphantom{\frac{r_{ij}^{\mu} r_{ij}^{\nu}}{r_{ij}^2}}
 \delta_{\mu \nu}
\left[ 1 - i k_0 r_{ij} - (k_0 r_{ij})^2 \right]
\right.
 \\
&-&\left. \frac{\mathbf{r}_{ij}^{\mu} \mathbf{r}_{ij}^{\nu}}{r_{ij}^2}
\left[3 - 3 i k_0 r_{ij} - (k_0 r_{ij})^2 \right]
\right\}.
\nonumber
\label{eq:green}
\end{eqnarray}
Here we assume that in the states $e$ and $e'$ atoms $i$ and $j$ are excited; $\mathbf{d}_{e g}$ is the matrix element of the dipole moment operator for the transition ${g} \to {e}$, $\mathbf{r}_{ij} =\mathbf{r}_i - \mathbf{r}_j$, $r_{ij} = |\mathbf{r}_i - \mathbf{r}_j|$ and
$k_0=\omega_0/c$ is the wavenumber associated to the transition, with $c$ the vacuum speed of light. The superscripts $\mu$ or $\nu$ denote projections of vectors on one of the axes $\mu$, $\nu = x$, $y$, $z$ of the reference frame.

In traditional approach to study of collective effects in cold atomic clouds, the system (\ref {e1}) is solved in steady state regime many times for various random spatial configurations of motionless atoms. In this paper, we consider moving atoms and search for nonstationary solution of Eq. (\ref{e1}). The displacement of atoms is given by the explicit dependence $ \mathbf {r} _i = \mathbf {r} _{i0} + \mathbf {v}_i t $. The distribution of atoms at  initial time $t=0$ is considered random, but spatially homogeneous on average. The velocities of the atoms at $t=0$ are also considered random variables. Their projections are assumed to be distributed according to the gaussian law.
\begin{equation}\label{3}
f(v_\mu)=1/\sqrt{2\pi v_0^2}\exp(-v_\mu^2/2v_0^2),
\end{equation}
Here $ v_\mu $ is one of the projections of the velocity; $ \mu = x, y, z $.
The dispersion of the velocities $ v_0 $ along three axes is assumed to be the same. The value of $ v_0 $ and the wave number $ k_0 $ determine the Doppler broadening of the line $ \Delta_D = 2 \sqrt {2 \ln2} k_0v_0 $. We will consider typical conditions for dipole traps, when the atomic temperatures are of the order of $30-100\; \mu K$. At such temperatures, the momentum of an atom with a mass of about 80 is several tens of times greater than the momentum of a photon. This allows us to consider the motion of atoms classically and not take into account the change in speed due to recoil.

In order not to take into account the possible withdrawal of atoms from the volume under consideration we assume that the volume of the cloud is surrounded by imagine surfaces on which the atoms scatter elastically. The system (\ref {e1}) under these assumptions is solved numerically.

From the computed values of $b_{e}(t)$, we can find the amplitudes of all other states which determine the wave function $\psi$ of the joined atom-field system (for more detail see \cite{KSH11}) and consequently the properties of the both atomic ensemble and the scattered light.

The first quantity we calculate here is the transmission of the plain layer of moving atoms. The intensity of light with polarization $\mathbf{u}$ ($|\mathbf{u}|$=1) transmitted through the sample can be written as a result of interference of incident and scattered waves:
\begin{eqnarray}
I(\mathbf{r}, \mathbf{u}, t) &=& \frac{c}{4\pi}
\left|
\mathbf{u}^* \cdot \mathbf{E}_{\mathrm{in}}(\mathbf{r}) \vphantom{\sum}
\right.
 \\
&+&  \left.
k_0^3 \sum\limits_{e}
f_{e}(\mathbf{r}, \mathbf{u}, t) b_e(t)
\vphantom{\sum}
\right|^{2},
\nonumber
\label{e2}
\end{eqnarray}
where
\begin{eqnarray}
f_{e}(\mathbf{r}, \mathbf{u}, t) &=&
\frac{e^{i k_0 |\mathbf{r} - \mathbf{r}_j|}}{k |\mathbf{r} - \mathbf{r}_j|}
\left[ \mathbf{u}^* \cdot \mathbf{d}_{g e}
\vphantom{\frac{[\mathbf{u}^* \cdot (\mathbf{r} - \mathbf{r}_j)]
[ \mathbf{d}_{g_j e_{jm}} \cdot (\mathbf{r} - \mathbf{r}_j)]}{|\mathbf{r} - \mathbf{r}_j|^2}}
\right.
 \\
&-& \left. \frac{[\mathbf{u}^* \cdot (\mathbf{r} - \mathbf{r}_j)]
[ \mathbf{d}_{g e} \cdot (\mathbf{r} - \mathbf{r}_j)]}{|\mathbf{r} - \mathbf{r}_j|^2}
\right].\;\;
\nonumber
\label{e3}
\end{eqnarray}
Here $\mathbf{E}_{\mathrm{in}}(\mathbf{r}) = \mathbf{u}_{\mathrm{in}} E_0 \exp(i \mathbf{k}_{\mathrm{in}} \mathbf{r})$; $\mathbf{k}_{\mathrm{in}}$ and  $\mathbf{u}_{\mathrm{in}}$ are the wave vector and the unit polarization vector of the incident light.

\begin{figure}\center
	\includegraphics[width=6cm]{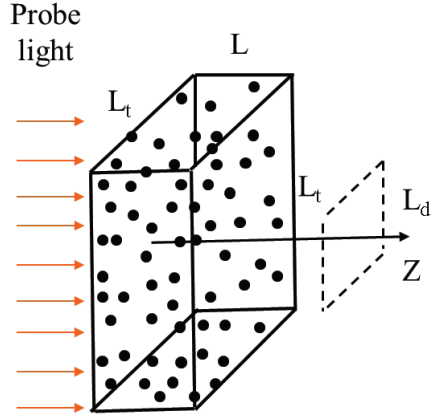}
	\caption{\label{fig:one}
			The considered model geometry. }\label{f1}
\end{figure}

Specific calculation is performed for the geometry shown in Fig. 1. To simplify the description of the reflection of atoms from boundaries, we consider an ensemble having the shape of a rectangular parallelepiped with transverse size $ L_t \times L_t $ and height $ L \ll L_t $. The finite transverse sizes of the sample lead to diffraction effects, and the spatial distribution of intensity (\ref {e2}) turns out to be non-uniform in the photo-detection plane. To be able to estimate the transmittance that would take place for  infinite slab, we calculate the signal averaged over the area of $ L_d \times L_d $ near the axis of the parallelepiped. In addition, taking into account the depolarization of the radiation as a result of multiple scattering in an optically dense medium, we calculate the total intensity of the transmitted light. Thus the studied transmittance can be calculated as follows
\begin{eqnarray}
 T(L, t)   = \frac{1}{I_0  L_d^2} \int\!\!\!\!\!\! \int\limits_{\!\!\!\!-L_d/2}^{\!\!\!\!L_d/2}\!\!dxdy  \int\limits_{4\pi} d^2\mathbf{u}\; I(\mathbf{r},  \mathbf{u}, t) ,\;\;\;\;\;\;\;
\label{e4}
\end{eqnarray}
where $I_0$ is the intensity obtained from Eqs. (\ref{e2}) and (\ref{e3}) in the absence of the atomic sample.

As it is known, the transmission coefficient is significantly affected by the presence of an external static electric or magnetic field \cite {Skipetrov_2019}. Thus, placing an atomic ensemble in a strong magnetic field induces a localization transition \cite {SS15}. For this reason, the transmission coefficient calculations will be carried out for this, the most important case. The magnetic field strength, as in \cite {SS15}, is considered to be so large that the Zeeman splitting $\Delta$ exceeds the typical shifts of the atomic levels caused by the dipole-dipole interaction.

In Fig. 2, as an example, we show the time dependence of the transmission (\ref {e4}) calculated for moving atoms with some arbitrarily chosen initial spatial and velocity distribution. The calculation was performed for a cloud with $ k_0L_t = 50 $ and $ k_0L = 6 $. The size of a photodetector is $ L_d = L_t / 2 $; its plane is separated from the back surface of the parallelepiped by $ k_0 (z-L) = 12$  (see. Fig. 1). The density of atoms is $ n \lambdabar ^ 3 = 0.2 $ ($\lambdabar= k_0^{-1} $). External coherent radiation is assumed to be circularly polarized, its detuning from the frequency of the resonant transition in a free atom is $ \delta = 0.5 \gamma $, which, according to the results of \cite {Skipetrov_2018}, corresponds to the spectral region of strong localization of light. The source of this radiation is switched on at $ t = 0 $ when all atoms are in the ground state. The speed of the atoms is $ k_0v_0 = 0.025 $; the total Doppler width of the transition line is approximately seventeen times smaller than the natural width. Zeeman shifts due to the external magnetic field were chosen a hundred times larger than the natural width of the lines. For comparison, in Fig. 2, the dotted line shows the result for the case of fixed atoms. This result is obtained by averaging over a large ($ 10 ^ 4 $) number of random spatial configurations.
\begin{figure}\center

	\includegraphics[width=8cm]{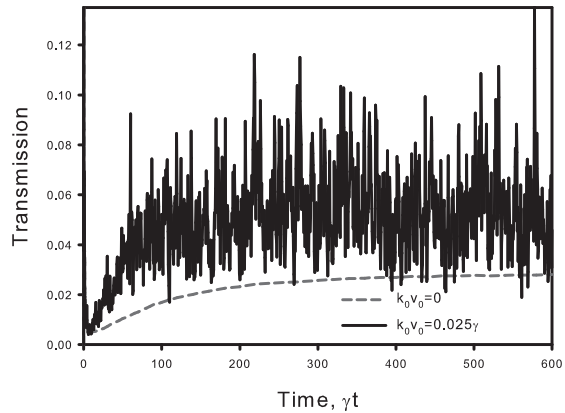}
	\caption{\label{fig:two}
			Comparison of the transmission dynamics of an atomic cloud for stationary (averaging over an ensemble of random spatial configurations) and moving atoms with some arbitrarily chosen initial velocity and space distributions. The calculation was performed for $k_0L_t=50$,  $k_0L=6$, $L_d=L_t/2$, $k_0(z-L)=12$ (see Fig. 1), $n\lambdabar^3=0.2$,  $\delta=0.5\gamma$, $\Delta=100\gamma$}\label{f2}
\end{figure}

The Fig. 2 clearly shows the three time scales of the change in the transmittance. The first is associated with the establishment of a quasi-stationary regime after switching on the source of external radiation. Characteristic times are approximately $ 150 \gamma ^ {- 1} $ for both moving and immobile atoms. This is the characteristic time for the population of the main part of collective excited states.

The second and the third scales are observed only for moving atoms. The transmittance experiences noticeable changes at times of the order of $ \gamma ^ {- 1} $.  Here we see a direct result of interference of secondary waves scattered by different atoms of the ensemble, which is very sensitive to the specific atomic configuration. Small change in this configuration leads to noticeable change in transmission. For motionless atoms, this interference manifests itself in a big difference of the results for different spatial configurations. And at last we can see the modification of transition at times approximately equal to $ 20-40\gamma ^ {- 1}$. It is the time it takes for the atoms to displace at $0.5-1 \lambdabar$ which modify their dipole-dipole interaction.

The second important result, which is shown in Fig. 2, is the difference between the time average calculated for moving atoms and the average over the ensemble of different spatial configurations for fixed atoms with the same density. The average transmittance for motionless atoms is approximately two times less, although the Doppler shifts at the considered velocities cannot lead to any noticeable modification of the dipole-dipole interaction.

Note that alteration of initial conditions for moving atoms essentially change the specific dependence $T(L,t)$. However our calculation shows that the time average transmission calculated for time interval $150<\gamma t<2000$ varies for considered parameters within no more than two percents of its value.  Taking into account this alteration further in calculation of quasi-stationary transmission $T(L)$ we make double averaging. First we calculate time average over time interval $\gamma \Delta t\simeq 1000$ for given initial conditions and then perform additional averaging over about one hundred initial conditions.

Accounting for the motion of atoms in the case of an ensemble in a magnetic field leads not only to quantitative but also to qualitative differences. Figure 3 shows the dependence of the quasi-stationary transmittance $T(L)$ on the thickness of the atomic ensemble for different average velocities of the atoms. Note that for all the velocities considered in Fig. 3, a change in the dipole-dipole interaction in comparision with motionless atoms is insignificant.

\begin{figure}\center

	\includegraphics[width=8cm]{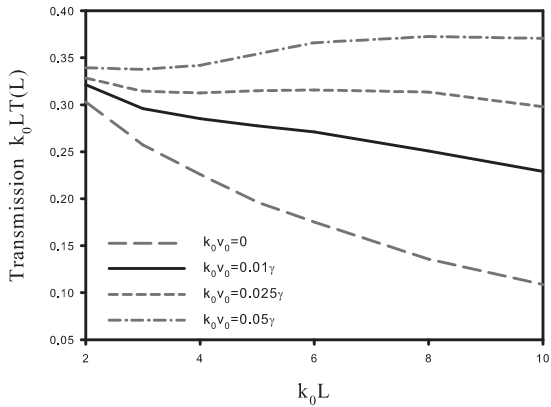}
	\caption{\label{fig:Three}
	Transmission coefficient of the atomic ensemble shown in the Fig. 1 multiplied by its thickness $k_0L$ for the radiation detuned at $\delta=0.5\gamma$ from the resonant frequency of the free atom. The calculation is performed for  $k_0L_t=50$, $L_d=L_t/2$, $k_0(z-L)=12$,$n\lambdabar^3=0.2$, $\Delta=100\gamma$.}\label{f3}
\end{figure}

It is seen that for immobile atoms ($ k_0v_0 = 0 $), the product of the transmittance by the thickness of the cloud is a fast decreasing function of the thickness. This corresponds to the strong localization of light predicted in \cite {SS15}. Heating leads to a significant change in the nature of dependence. The transition to the diffuse transfer regime is clearly visible. Even at temperatures corresponding to $ k_0v_0 = 0.025 $, when the Doppler broadening is seventeen times less than the natural linewidth, we do not observe an exponential decrease in transmission (see, for example, \cite {SS19}). At speeds of $ k_0v_0 = 0.01 $, when the $ \gamma/\Delta_D \sim 40 $, we see some manifestations of localization, but the localization length, determined by the rate of decrease in transmission, turns out to be significantly larger than for motionless atoms.

The observed effects, in our opinion, can be explained by the change in the spectrum of collective states of an ensemble with time caused by atomic displacement. Among the collective states there are long-lived, sub-radiant. For their excitation  a long-term exposure by external radiation is required. The excitation time should be several times larger than corresponding lifetime. But at such time intervals the motion of atoms can significantly change the spectrum of states. The time of effective excitation of each given state is always finite and is determined by the speed of atoms. And, while for super-radiant states this time for considered temperatures is enough, for long-lived it is not the case. In the ensemble of moving atoms, there is always a group of long-lived states with narrow energy levels that simply do not have time to be effectively populated. The higher the speed of motion of atoms, the more such states. And when replacing an ensemble of slowly moving atoms with a stationary ones in traditional approaches, we do not take into account this effect, which actually turns out to be significant.

The difference in influence of atomic motion on sub and super-radiant states are most clearly manifested in the dynamics of spontaneous decay, see Fig. 4. We considered spontaneous decay in a polyatomic ensemble, in which one of the atoms located in the center of a model cubic cloud was initially excited on a certain Zeeman sublevel. With time, the population of this state (we denote it by the index $ s $), as well as all the others, changes. To calculate the population dynamics, we solve the system (\ref {e2}) with a given initial condition and in the absence of external radiation. In theory, we can retrace any of the populations, including the total population of all excited states. As an example in Fig. 4 we show the time dependence of  population of the initially excited state $ s $, i.e. the dependence $ P_s = | b_s (t) | ^ 2 $. To demonstrate that the movement of atoms noticeably affects collective phenomena even in the case when there are no anomalously long-lived states, we considered the ensemble in the absence of external static fields. As is well known (\cite {SS14, BGAK14}), localized long-lived states are absent in such ensembles at any density.
\begin{figure}\center
	\includegraphics[width=8cm]{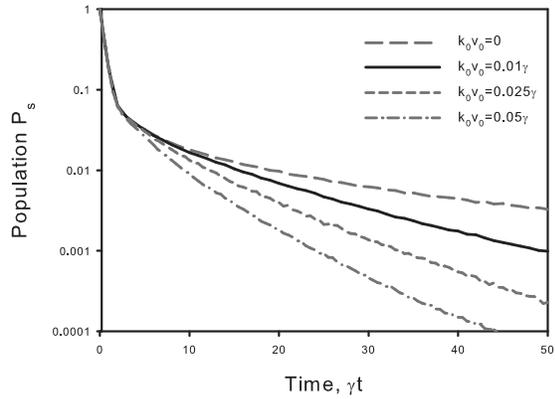}
	\caption{\label{fig:Four}
			Spontaneous decay dinamics of the atom excited in the center of the atomic cloud; $k_0L_t=k_0L=16$, $n\lambdabar^3=0.1$.}\label{f4}
\end{figure}

The calculations of time dependence $ P_s = | b_s (t) | ^ 2 $ were carried out for different atomic velocities for $ k_0L_t = k_0L = 16 $. The density of atoms is $ n \lambdabar ^ 3 = 0.1 $. Since spontaneous decay is a substantially non-stationary process, to obtain the curves in Fig. 4, we averaged over the results of multiple calculations of the dynamics for different initial spatial configurations and different initial velocities of atoms.

An increase in the velocity of atoms has practically no effect on the decay rate at small times. Herewith, at large times, in which long-lived states have a major influence on the dynamics of the system, the decay rate increases with increasing in $ v_0 $. Even at $ k_0v_0 = 0.01 $ and $ \gamma t \sim 40 $, the instantaneous decay rate is two times higher than that observed for fixed atoms. The higher the atomic velocities, the less long-lived collective states participate in the formation of the dynamics of the system at large times.

In conclusion, we considered the influence of atomic motion on the character of collective effects in dense and cold atomic ensembles. We have shown that even in the case when the characteristic Doppler shifts are several tens times smaller than the natural width of atomic excited states, the displacement of atoms significantly suppresses the role of long-lived, sub-radiant collective states. This suppression leads to noticeable quantitative changes in the behavior of the systems under consideration. It can also lead to qualitative changes, for example, to a transition from the strong localization regime of radiation transfer to diffusion one.

\section*{Acknowledgments}
This work was supported by the Russian Science Foundation (Grant No. 17-12-01085).
The results of the work were obtained using computational
resources of Peter the Great Saint-Petersburg Polytechnic
University Supercomputer Center (http://www.scc.spbstu.ru).

\end{document}